\def \SAIT #1 #2 {{\em Mem.\ Soc.\ Astron.\ It.\/} {\bf #1}, #2}
\def \MESS #1 #2 {{\em The Messenger\/} {\bf #1}, #2}
\def \ASTRNACH #1 #2 {{\em Astron. Nach.\/} {\bf #1}, #2}
\def \AAP #1 #2 {{\em Astron. Astrophys.\/} {\bf #1}, #2}
\def \AAL #1 #2 {{\em Astron. Astrophys. Lett.\/} {\bf #1}, L#2}
\def \AAR #1 #2 {{\em Astron. Astrophys. Rev.\/} {\bf #1}, #2}
\def \AAS #1 #2 {{\em Astron. Astrophys. Suppl. Ser.\/} {\bf #1}, #2}
\def \AJ #1 #2 {{\em Astron. J.\/} {\bf #1}, #2}
\def \ANNREV #1 #2 {{\em Ann. Rev. Astron. Astrophys.\/} {\bf #1}, #2}
\def \APJ #1 #2 {{\em Astrophys. J.\/} {\bf #1}, #2}
\def \APJL #1 #2 {{\em Astrophys. J. Lett.\/} {\bf #1}, L#2}
\def \APJS #1 #2 {{\em Astrophys. J. Suppl.\/} {\bf #1}, #2}
\def \APSS #1 #2 {{\em Astrophys. Space Sci.\/} {\bf #1}, #2}
\def \ASR #1 #2 {{\em Adv. Space Res.\/} {\bf #1}, #2}
\def \BAIC #1 #2 {{\em Bull. Astron. Inst. Czechosl.\/} {\bf #1}, #2}
\def \JSQRT #1 #2 {{\em J. Quant. Spectrosc. Radiat. Transfer\/} {\bf #1}, #2}
\def \MN #1 #2 {{\em Mon. Not. R. Astr. Soc.\/} {\bf #1}, #2}
\def \MEM #1 #2 {{\em Mem. R. Astr. Soc.\/} {\bf #1}, #2}
\def \PLR #1 #2 {{\em Phys. Lett. Rev.\/} {\bf #1}, #2}
\def \PASJ #1 #2 {{\em Publ. Astron. Soc. Japan\/} {\bf #1}, #2}
\def \PASP #1 #2 {{\em Publ. Astr. Soc. Pacific\/} {\bf #1}, #2}
\def \NAT #1 #2 {{\em Nature\/} {\bf #1}, #2}

\documentstyle[twoside]{memsait}
\begin{opening}
\title{THE X--RAY SPECTRUM OF COMPTON--THICK SEYFERT 2 GALAXIES}
\author{GIORGIO MATT}
\institute{Dipartimento di Fisica, Universit\`a degli Studi ``Roma Tre",
Via della Vasca Navale 84, I--00146 Roma, Italy}
\date{} 
\end{opening}

\begin{document}

\oddpagefooter{}{}{} 
\evenpagefooter{}{}{} 
\ 
\bigskip

\begin{abstract}
Current ideas on the X--ray spectrum of Compton--thick Seyfert
2 galaxies are reviewed, and a brief description of the four presently
known sources of this class are given. 
\end{abstract}

\section{Introduction}

In the 
unified model for Seyfert galaxies (Antonucci 1993 and references
therein), type 2 sources are believed to be intrinsically identical to 
type 1's, but observed at inclination angles, with respect
to the symmetry axis of the molecular torus (see e.g. Ward
1994), greater than the torus 
half--opening angle. Therefore, at least in optical/UV, the nucleus
of Seyfert 2 galaxies is not directly visible.

In X--rays the situation is more complex. In this band, the two most
important interactions between photons and (cold) matter are photoabsorption
and Compton scattering. The photoabsorption cross section
strongly depends on energy (decreasing approximately as $E^{-3.5}$), 
while the Compton cross section is constant, at least up to 
the Klein--Nishina decline. The two cross sections, for solar
chemical composition, are equal at about 10 keV. 
If the column density of the
torus is smaller than $\sigma_{\rm T}^{-1}\sim1.5\times 10^{24}$ 
cm$^{-2}$, the nucleus turns out to
be visible above a few keV, and the sources are named 
{\it Compton--thin}, because the matter is optically thin to
Compton scattering. If, on the contrary, the column density exceeds that 
value, the matter is optically thick to Compton scattering; for
Compton optical depths of $\sim$a few, the nucleus becomes practically
invisible also in hard X--rays because, after a few scatterings, photons 
are redshifted down to the photoabsorption
dominated regime. These sources are called {\it Compton--thick} Seyfert 2
galaxies, and can be observed in X--rays only in scattered light, as will be
discussed in the next section. Of course, they
are very faint in X--rays and only a handful (four, excluding the borderline
source NGC 4945) of them are presently known.

\section{The X--ray spectrum of Compton--thick sources}

Even if the type-1 nucleus is, in Compton--thick
sources, completely obscured in the whole X--ray band,
its existence is revealed by scattered light. Two 
scattering media are
thought to be present in the vicinity of the nucleus: 
$(a)$ the optically thin matter responsible for
the scattering and polarization of the optical 
broad lines in several Seyfert 2's (the so--called ``warm mirror"),
and $(b)$ the inner surface of the torus itself which, in these sources, 
is optically thick by definition.

\noindent
$(a)$ Scattering from the warm mirror of the nuclear continuum radiation 
produces a spectrum which is practically identical to the nuclear
spectrum up to energies at which Compton recoil is important, 
where a cut--off occurs (Matt 1996; Poutanen et al. 1996). 
Therefore, its shape is expected to be, at least below say a few tens of keV, 
a power law with photon index $\sim$2 (Nandra \& Pounds 1994).
Line emission from ionized atoms is also expected, iron K$\alpha$ lines being
among the most prominent (other lines, e.g. from carbon, oxygen and
neon, can also be very intense; however, they are often diluted
by other continua arising in soft X--rays, e.g. from starburst regions. 
Fe {\sc xxv} (6.7 keV) and Fe {\sc xxvi} 
(6.97 keV) K$\alpha$ line emission from such a medium is
discussed in detail by Matt, Brandt \& Fabian (1996). Their equivalent
width with respect to the continuum reflected
from the warm mirror itself can be as high as a few keV. 

\noindent
$(b)$ As shown by Ghisellini, Haardt \& Matt (1994) and Krolik, Madau \&
\.Zycki (1994), 
the spectrum of the nuclear radiation scattered from the inner surface of the
torus is similar to
that observed in many Seyfert 1 galaxies (Mushotzky, Done \& Pounds 
1993 and references therein), where the reflecting matter is 
believed to be the inner accretion disc. It is usually referred
to as the Compton reflection component. It is very flat in the classical
2--10 keV band (photon index less than 1) due to the increasing ratio of
Compton scattering to photoabsorption cross sections (e.g. 
Lightman \& White 1988). A fluorescent K$\alpha$ line at 6.4 keV from
neutral iron, 
with an equivalent width with respect to the continuum reflected by 
the same matter 
of about 1--2 keV (e.g. Matt et al. 1991; Ghisellini, Haardt \& Matt 1994;
Reynolds et al. 1994)
is also expected, as well as fainter lines from lighter elements
(Reynolds et al. 1994; Matt, Fabian \& Reynolds 1996).

The intensity of the radiation from the warm mirror 
depends mainly on the optical
depth of the mirror itself, while that of the torus component depends
basically on the inclination angle of the system (see Matt, Brandt \& Fabian 
1996). Their 
relative ratio can therefore be very different from source to source. Three
cases are then possible:

\noindent
{\sc i}) The warm mirror component dominates over the torus component. A 
2--10 keV spectrum
composed by a power law with photon index about 2 plus strong ionized
lines is expected. No sources of this kind have been discovered
yet to our knowledge.

\noindent
{\sc ii}) The two components are of the same order. A spectrum with an 
intermediate or flat 
photon index (1--1.5 or less) and both neutral and ionized lines is expected, and
it has indeed been observed in NGC~1068 (Ueno et al. 1994; see below for
a revised analysis) and NGC~6240 (Iwasawa, private communication).
It is worth stressing that it is possible to observe in one and the same source 
a continuum dominated, above a few keV, by the torus component {\sl and}
also a complex iron line, as the lines from ionized matter can be
very bright and then visible even if the relative continuum is much smaller
than the torus one (this actually seems to be the case for the two sources of 
this group). 

\noindent
{\sc iii}) The torus component dominates over the warm mirror component. 
A very flat (less than one)
spectrum with a strong 6.4 keV iron line is expected, and actually
observed in NGC~6552 (Fukazawa et al. 1994) and in the Circinus 
Galaxy (Matt et al. 1996). 

In the next section we will briefly discuss each of the four currently
known Compton--thick sources. 

\section{Group II sources}

{\bf NGC~1068} is surely the best studied among Compton--thick Seyfert 2's 
(and maybe among all Seyfert 2's). Its X--ray spectrum is composed
by thermal--like emission, probably due to a starburst region (Wilson et al. 
1994) and dominating below 3--4 keV, and a hard power law with superposed
a complex iron line, composed by a neutral component at 6.4 keV and
contributions from He-- and H--like iron (Ueno et al. 1994 and references
therein). Iwasawa, Fabian \& Matt (1996) have recently re--analysed the ASCA PV
observation and found that: a) the photon power law index is smaller than
previously estimated, being less than 0.4. This value is consistent with a
pure reflection continuum. 
 b) A feature just redwards of the 
neutral iron line is present, with a flux of about 10 percent of that of 
the line, and 
interpreted as the line Compton shoulder, i.e. line photons scattered
once before escaping from the matter. This feature is predicted by the 
reflection models (see e.g. Matt, Perola \& Piro 1991) but never detected so far.
c) The ionized iron lines
are both redshifted; the derived velocity of the emitting matter is 
about 4000--5000 km/s. 

The interpretation of the spectrum above a few keV is that we are
looking at reflection from both the warm mirror and the obscuring torus, 
with the former component smaller but not completely overwhelmed by the latter
one, so that the continuum is dominated by the torus reflection but 
the lines from the ionized matter (possibly forming a wind,
as suggested by its relatively high velocity) are still visible.


\noindent
{\bf NGC~6240} is one of the most famous merging system and a 
prototype ultraluminous infrared galaxy.
In X--rays, the source has been observed by both ROSAT--PSPC (Fricke \& 
Papaderos 1996) and ASCA (K. Iwasawa, private communication).
The ROSAT--PSPC revealed extended soft 
X--ray emission, which can be explained by a starburst model. 
The ASCA spectrum is complex, consisting of: a thermal--like component,
consistent with the ROSAT observation and dominating in soft X--rays; 
a quite hard power law;
both neutral and ionized iron lines. The hard spectrum is 
very similar to that of NGC~1068, and can be explained in the same way.

\section{Group III sources}

{\bf NGC 6552} was serendipitously observed by ASCA 
(mainly thanks to the very prominent
6.4 keV iron line; Fukuzawa et al. 1994), and then 
identified with a Seyfert 2 galaxy.
Its X--ray spectrum is 
well described by a pure Compton reflection continuum (Reynolds et 
al. 1994), and then interpreted as reflection from cold matter (possibly
the torus) of an otherwise invisible X--ray nucleus. A heavily absorbed
power law maybe also be present, but the source is too faint 
to permit any strong conclusion on this and other spectral details.

\noindent
The {\bf Circinus Galaxy} is a nearby (4 Mpc) Seyfert 2 with
a very prominent
ionization cone, strong coronal lines and water maser emission (see Oliva
et al. 1994 and references therein). Before ASCA, it was
never observed in X--rays, apart from a detection in the ROSAT all sky survey.
The ASCA spectrum is quite remarkable, showing a very prominent iron 
K$\alpha$ line
at 6.4 keV, many other lines from lighter elements as well as the K$\beta$
iron line, and a very hard continuum (Matt et al. 1996). 
Below about 2 keV, a further component
emerges, possibly due to a starburst region. Above that energy, the spectrum 
is well described by a pure reflection component. Lines from 
elements lighter than iron 
are also expected in this model (Reynolds et al. 1994;
Matt, Fabian \& Reynolds 1996). However, their 
observed intensities and centroid energy
are not entirely consistent with reflection by cold matter; improved 
ASCA--SIS calibration matrices have to be waited for 
before further addressing this issue.

\section{Conclusions}

The model described in Sec.2 for the X--ray spectrum of Compton--thick 
Seyfert 2 galaxies is fully consistent with the observations
of the four known sources of this class. A prediction of this model
is a spectral cut--off at a few tens of keV due to Compton recoil.
BeppoSAX and RXTE observations will then be valuable in testing the model.

\acknowledgements
I thank all the colleagues who collaborate with me on this topic. 
I'm particularly indebted to K. Iwasawa for allowing me to quote
his results on NGC 6240 before publication.


\end{document}